\documentstyle[12pt,epsf]{article}   
\begin{document}    
\baselineskip=24.5pt    
\setcounter{page}{1}         
\topskip 0 cm    
\vspace{1 cm}    
\centerline{\bf Predictions from the Fritzsch-Type Lepton Mass Matrices}
\vskip 1 cm
\centerline{M. Fukugita$^{1}$, M. Tanimoto$^2$ and T. Yanagida$^3$}
\vskip5mm
\centerline{$^1$ Institute for Cosmic Ray Research, University of Tokyo,
Kashiwa 277 8582, Japan}
\centerline{$^2$ Department of Physics, Niigata University, Niigata 950 2181,
Japan}
\centerline{$^3$ Department of Physics, University of Tokyo,  Tokyo 113 0033, 
Japan}

\vskip 4 cm
\noindent
{\bf Abstract}

We revisit the Fritzsch-type lepton mass matrix models confronted with
new experiments for neutrino mixings. It is shown that the model is
viable and leads to a rather narrow range of free parameters.
Using empirical mixing information between  $\nu_e$ and $\nu_\mu$,
and between  $\nu_\mu$ and $\nu_\tau$, 
it is predicted that the mixing angle between
$\nu_e$ and $\nu_\tau$ is in the range $0.04<|U_{13}|<0.20$,
consistent with the CHOOZ experiment and the lightest neutrino mass is
$0.0004<m_1<0.0030$ eV. The range of the effective mass measured in
double beta decay is $0.002<\langle m_{ee}\rangle< 0.007$ eV. 

\newpage

%

During the last five years we have experienced dramatic advancement in 
empirical understanding of the mass and mixing of neutrinos.
Most recently, the KamLAND experiment selected the neutrino
mixing solution that is responsible for the solar neutrino problem
nearly uniquely [1]: we are left with only so-called 
`large-mixing angle solution'
(LMA).
We have now good understanding concerning the neutrino
mass difference squared and mixing between $\nu_e$ and $\nu_\mu$ [2],
and between $\nu_\mu$  and $\nu_\tau$ [3]. An interesting constraint has also
been placed on mixing between  $\nu_e$ and $\nu_\tau$ from the reactor
experiment [4].

Anticipating that neutrinos are all massive from an early indication of
the Kamiokande experiment [5], we  proposed [6] that neutrino
mass and mixing may be described by mass matrices similar to those
proposed by Fritzsch for quarks [7]. After the secure confirmation of
neutrino oscillation at SuperKamiokande [3], a large number of mass
matrix models are proposed [8], but many of them among interesting models
lead to bimaximal mixing.
We now know that mixing between $\nu_e$ and $\nu_\mu$ is large but
not maximal [2], whereas mixing between  $\nu_\mu$ and $\nu_\tau$
may be maximal [3]. This mixing pattern, together with small mixing between
$\nu_e$ and $\nu_\tau$, is  a characteristic embedded in
the Fritzsch type matrices we proposed.

So, we revisit the Fritzsch-type mass matrix model to examine whether it
is still consistent with all experiments. We find that it is, but
the model leads, in particular after the KamLAND experiment, 
to a narrow allowed parameter range that can be tested in neutrino 
experiments we expect in the future.

A similar analysis was recently done by Xing [9]. The author, however,
fixed the phases that appear in matrices to some arbitrary values. 
So his result is only 
a limited representation of the model. In this paper we have 
exhaustively studied for the allowed parameter space of the model.


The model we proposed in [6] consists of 
the mass matrices of the charged leptons and the Dirac neutrinos
of the form
\begin{eqnarray}
m_E = \left( \matrix{0 & A_\ell & 0 \cr A_\ell & 0 & B_\ell \cr
                     0 & B_\ell & C_\ell\cr        } \right)\ ,\qquad\quad
m_{\nu D} = \left( \matrix{0 & A_\nu & 0 \cr A_\nu & 0 & B_\nu \cr
                     0 & B_\nu & C_\nu\cr        } \right)\  , 
\end{eqnarray}
where each entry is complex, and the right-handed Majorana mass matrix 
\begin{eqnarray}
M_R = M_0 {\bf I},
\end{eqnarray}
where $M_0$ is a very large mass. We assume that neutrinos are of the
Majorana type.
We obtain small neutrino masses $m_1$,  $m_2$ and $m_3$
through the seesaw mechanism, as
\begin{equation}
m_i=m_{\nu D}^TM_R^{-1}m_{\nu D}.
\end{equation}
The lepton mixing matrix is 
given by\footnote{Note that in [6] the mixing matrix is
defined by $U=U_\nu^\dagger Q U_\ell$ in analogy 
to that in the quark sector. So $U_{ij}$ is replaced with $U_{ji}^*$,
when the present paper is compared with [6].}
\begin{eqnarray}
 U = U_\ell^\dagger \ Q \ U_\nu
\end{eqnarray}
where
\begin{eqnarray}
&&U_{\ell}(1,1)=\sqrt{\frac{m_{2e}m_{3e}(m_{3e}-m_{2e})}{(m_{2e}+m_{1e})
                               (m_{3e}-m_{2e}+m_{1e})(m_{3e}-m_{1e})}}
\nonumber\\
&&U_{\ell}(1,2)=-\sqrt{\frac{m_{1e}m_{3e}(m_{3e}+m_{1e})}{(m_{2e}+m_{1e})
                               (m_{3e}-m_{2e}+m_{1e})(m_{3e}+m_{2e})}}
\nonumber\\
&&U_{\ell}(1,3)=\sqrt{\frac{m_{1e}m_{2e}(m_{2e}-m_{1e})}{(m_{3e}-m_{1e})
                            (m_{3e}-m_{2e}+m_{1e})(m_{3e}+m_{2e})}}
\nonumber\\
&&U_{\ell}(2,1)=\sqrt{\frac{m_{1e}(m_{3e}-m_{2e})}{(m_{2e}+m_{1e})(m_{3e}-m_{1e})}}
\nonumber\\
&&U_{\ell}(2,2)=\sqrt{\frac{m_{2e}(m_{3e}+m_{1e})}{(m_{2e}+m_{1e})(m_{3e}+m_{2e})}}
\nonumber\\
&&U_{\ell}(2,3)=\sqrt{\frac{m_{3e}(m_{2e}-m_{1e})}{(m_{3e}+m_{2e})(m_{3e}-m_{1e})}}
\nonumber\\
&&U_{\ell}(3,1)=-\sqrt{\frac{m_{1e}(m_{2e}-m_{1e})(m_{3e}+m_{1e})}{(m_{3e}-m_{1e})(m_{3e}-m_{2e}+m_{1e})(m_{2e}+m_{1e})}}
\nonumber\\
&&U_{\ell}(3,2)=-\sqrt{\frac{m_{2e}(m_{2e}-m_{1e})(m_{3e}-m_{2e})}{(m_{3e}+m_{2e}) (m_{3e}-m_{2e}+m_{1e})(m_{2e}+m_{1e})}}
\nonumber\\
&&U_{\ell}(3,3)=\sqrt{\frac{m_{3e}(m_{3e}+m_{1e})(m_{3e}-m_{2e})}{(m_{3e}+m_{2e})(m_{3e}-m_{2e}+m_{1e})(m_{3e}-m_{1e})}} \nonumber\\
\end{eqnarray}
\begin{eqnarray}
&&U_{\nu}(1,1)=\sqrt{\frac{m_{2D}m_{3D}(m_{3D}-m_{2D})}{(m_{2D}+m_{1D})
                               (m_{3D}-m_{2D}+m_{1D})(m_{3D}-m_{1D})}}
\nonumber\\
&&U_{\nu}(1,2)=-\sqrt{\frac{m_{1D}m_{3D}(m_{3D}+m_{1D})}{(m_{2D}+m_{1D})
                               (m_{3D}-m_{2D}+m_{1D})(m_{3D}+m_{2D})}}
\nonumber\\
&&U_{\nu}(1,3)=\sqrt{\frac{m_{1D}m_{2D}(m_{2D}-m_{1D})}{(m_{3D}-m_{1D})
                            (m_{3D}-m_{2D}+m_{1D})(m_{3D}+m_{2D})}}
\nonumber\\
&&U_{\nu}(2,1)=\sqrt{\frac{m_{1D}(m_{3D}-m_{2D})}{(m_{2D}+m_{1D})(m_{3D}-m_{1D})}}
\nonumber\\
&&U_{\nu}(2,2)=\sqrt{\frac{m_{2D}(m_{3D}+m_{1D})}{(m_{2D}+m_{1D})(m_{3D}+m_{2D})}}
\nonumber\\
&&U_{\nu}(2,3)=\sqrt{\frac{m_{3D}(m_{2D}-m_{1D})}{(m_{3D}+m_{2D})(m_{3D}-m_{1D})}}
\nonumber\\
&&U_{\nu}(3,1)=-\sqrt{\frac{m_{1D}(m_{2D}-m_{1D})(m_{3D}+m_{1D})}{(m_{3D}-m_{1D})(m_{3D}-m_{2D}+m_{1D})(m_{2D}+m_{1D})}}
\nonumber\\
&&U_{\nu}(3,2)=-\sqrt{\frac{m_{2D}(m_{2D}-m_{1D})(m_{3D}-m_{2D})}{(m_{3D}+m_{2D}) (m_{3D}-m_{2D}+m_{1D})(m_{2D}+m_{1D})}}
\nonumber\\
&&U_{\nu}(3,3)=\sqrt{\frac{m_{3D}(m_{3D}+m_{1D})(m_{3D}-m_{2D})}{(m_{3D}+m_{2D})(m_{3D}-m_{2D}+m_{1D})(m_{3D}-m_{1D})}} \nonumber\\
\end{eqnarray}
where 1-3 refers to generation; $Q$ is a phase matrix
written as

\begin{eqnarray}
Q= \left(\matrix{ 1 & 0 & 0 \cr 0 & e^{i \sigma} & 0\cr  0 & 0 & e^{i \tau}
\cr } \right)\ ,
\end{eqnarray}
which is a
reflection of phases contained in the 
charged lepton mass matrix and the Dirac mass matrix of neutrinos.
The phases are neglected in the right-handed neutrinos mass matrix.
In the presence of mass hierarchy of the left-handed neutrinos,
the effect of phases in the right-handed mass matrix is small,
and the inclusion of phases changes the anlysis only little.
Since the charged lepton masses are well-known, the
number of parameters  contained in our model is six: 
$m_{1D},\ m_{2D}, \ m_{3D}$, $\sigma$, $\tau$ and $M_0$.
These parameters are to be determined by empirical
neutrino mass and mixing.

We calculate lepton mixing matrix elements using the exact 
expressions, but they are written approximately

\begin{eqnarray}
U_{12} &\simeq& -\left({m_1 \over m_2}\right)^{1/4}+
\left({m_e\over m_\mu}\right)^{1/2}e^{i\sigma} \cr
U_{21} &\simeq& \left({m_1\over m_2}\right)^{1/4}e^{i\sigma}
-\left({m_e\over m_\mu}\right)^{1/2} \cr
U_{23} &\simeq&  \left({m_2\over m_3}\right)^{1/4}e^{i\sigma}
-\left({m_\mu\over m_\tau}\right)^{1/2} e^{i\tau}\cr 
U_{32} &\simeq&  -\left({m_2\over m_3}\right)^{1/4}e^{i\tau}
+\left({m_\mu\over m_\tau}\right)^{1/2} e^{i\sigma}\cr
U_{13} &\simeq&  \left({m_e\over m_\mu}\right)^{1/2}U_{23}+ 
\left({m_2\over m_3}\right)^{1/2} \left({m_1\over m_3}\right)^{1/4}
   \cr
U_{31} &\simeq&  \left({m_1\over m_2}\right)^{1/4}U_{32} 
\end{eqnarray}
where $(m_e/m_\tau)^{1/2}$ is  neglected.
Rough characteristics of mixing angles are understood from these expressions.


For our anlysis we take as inputs 

\begin{eqnarray}
\Delta m^2_{\rm atm}= (1.5 - 3.9) \times 10^{-3} {\rm eV^2}, 
\hskip5mm  \sin^2 \theta_{\rm atm}\geq 0.92
\end{eqnarray}
for $\nu_\mu-\nu_\tau$ mixing from Super-Kamiokande experiment [3],
and 

\begin{eqnarray}
\Delta m^2_{\rm sol}=(6 - 8.5)\times 10^{-5} {\rm eV}^2 , \hskip5mm
\sin^2 \theta_{\rm sol}=0.25 - 0.4
\label{eq:KamLAND-A}
\end{eqnarray}
for $\nu_e-\nu_\mu$ mixing from KamLAND [1], both at a 90\% confidence
level.

The KamLAND experiment gives another, but less favoured solution within LMA,
i.e., the solution allowed only at a 95\% confidence level,

 \begin{eqnarray}
\Delta m^2_{\rm sol}=(1.4 - 1.8)\times 10^{-4} {\rm eV}^2 , \hskip5mm
\sin^2 \theta_{\rm sol}=0.27 - 0.34
\end{eqnarray}
which we call KamLAND-B (we call the best favoured solution 
(\ref{eq:KamLAND-A}) KamLAND-A). For this case we take a region 
allowed at a 95\% confidence for the Super-Kamiokande data
for $\nu_\mu-\nu_\tau$ mixing 
for consistency. 

We discuss how the allowed parameter region shrinked after the KamLAND 
data which determined the mass difference squared to a high accuracy. For a
purpose of comparison, we also take the LMA for the $\nu_e-\nu_\mu$ mixing
before the KamLAND data [2]: 

\begin{eqnarray}
\Delta m^2_{\rm sol}=(3 - 20)\times 10^{-5} {\rm eV}^2 , \hskip5mm
\sin^2 \theta_{\rm sol}=0.23 - 0.41
\end{eqnarray}
at a 90\% confidence level.





We assume that $m_{3\nu}\gg m_{i\nu}$ for $i=1,2$, i.e., 
$\Delta m_{\ atm}^2\simeq m_{3}^2$. We do not assume
mass hierarchy between $m_{\nu 1}$ and $m_{\nu 2}$.
We do {\it not} include the empirical constraint from the CHOOZ 
experiment for mixing between $\nu_e$ and $\nu_\tau$ in our 
analysis. We leave this as
  a free parameter and examine the result against experiment.


We have exhaustively searched for allowed parameter regions 
that satisfy empirical mass difference squared and
mixing in six dimensional space without any prior
conditions. 

The result is restrictive. We plot in Figure 1 $m_1/m_3$ as
a function of $|U_{e3}|$ for KamLAND-A. 
The outer contour shows an allowed region 
before the KamLAND data are available. 
The model predicts  $0.04<|U_{e3}|<0.18$. It is
interesting to note that the model upper limit agrees
with the empirical constraint from the CHOOZ experiment
$|U_{e3}|<0.16$
and that there is a definite lower limit on $|U_{e3}|$.
The lightest neutrino mass $m_1$ cannot
be too small. When $m_2$ is used as a unit, 
$m_1/m_2= 0.05-0.29$ is the allowed range, or $m_1=0.0004-0.0030$ eV.
Of course, $m_1<m_2$ is always satisfied because $\theta_{\rm solar}$
does not reach 45$^\circ$. 

A similar figure is presented for KamLAND-B solution at
a 95\% confidence level (Figure 2).

The allowed range of the lepton flavour mixing matrix 
(for KamLAND-A) is given as

\begin{equation}
|U|= \left[ \matrix{0.76-0.86 & 0.50-0.63 & 0.04-0.19 \cr
                  0.27-0.48 & 0.63-0.72 & 0.60-0.71\cr
                  0.34-0.49  & 0.42-0.58 & 0.71-0.80 \cr
                                         } \right]\ .
\end{equation}
This matrix may be compared with 
a model-independent analysis of neutrino mixing [10],
which is modified very little even after the KamLAND experiment.
We see general agreement between the two matrices, while
the allowed ranges of each matrix element in the present 
model is quite narrow.   

In Figure 3  (Figure 4 for KamLAND-B) 
we show the rephasing invariant CP violation measure 
$J_{CP}$ [11], which is defined by
\begin{equation}
    J_{CP}= {\rm Im} \ [ {U_{\mu 3} U^*_{\tau 3}  U^*_{\mu 2} U_{\tau 2}} ]\ .
\end{equation}
as a function of $|U_{e3}|$.  

Figures 5 and 6 present experimentally more relevant quantities,
the CP violating part of neutrino oscillation 
 \begin{eqnarray}        
  \Delta P &\equiv& 
P(\overline \nu_\mu \rightarrow \overline \nu_e) - P(\nu_\mu \rightarrow\nu_e) =
P(\nu_\mu\rightarrow\nu_\tau)- P(\overline\nu_\mu\rightarrow\overline\nu_\tau)
  \nonumber \\
&=& P(\overline\nu_e\rightarrow\overline\nu_\tau)-P(\nu_e\rightarrow\nu_\tau)\ ,
   \label{CP}
   \end{eqnarray}
  \noindent
and effective mass measured in double beta decay experiment
$\langle m_{ee}\rangle$
\begin{equation}
 \langle m_{ee}\rangle = m_1\  U_{e1}^2 +  m_2\  U_{e1}^2 +  m_3\  U_{e3}^2
\end{equation}
for KamLAND-A case. Here, $\Delta P$ is given by

\begin{equation}
\Delta P=4 J_{CP} f_{CP}  
\end{equation}
where   
   \begin{equation}
f_{CP}\equiv -4\sin\frac{\Delta m^2_{21}L}{4E}\sin\frac{\Delta m^2_{32}L}{4E}            \sin\frac{\Delta m^2_{31}L}{4E}  \ ,
 \label{SCP}
   \end{equation}
   \noindent with 
$\Delta m_{ij}= m_i^2 - m_j^2$
and $\Delta P$ is evaluated for 
$\langle E_\nu \rangle =1.3 {\rm GeV}$ and $L=295 {\rm Km}$,
which are parameters for a planned long-baseline 
neutrino oscillation experiment
between the Japanese Hadron Facility (JHF)
in Tokai Village (Ibaraki) to the Super-Kamiokande. 

Note that the figure of $\langle m_{ee}\rangle$ doe not include
the error of $m_3$, but fixed at $0.053$ eV.  The result should be scaled 
if actual $m_3$ is higher or lower.
The effective mass measured in the double beta decay ranges from
$2-7$ meV for $m_3=0.053$ (eV). 
The lower limit is raised by a factor of 4 after the
KamLAND experiment.

Predictions are summarized in Table 1.
We emphasize that interesting features of this model are mixing
between $\nu_e$ and $\nu_\mu$ that is not maximal, unlike in the model 
proposed in [12] based on the democratic principle or that in
the Zee model [13,14], whereas $|U_{13}|$ is automatically predicetd to be
small (but non-zero) [8]. Basic features found in experiment seem to be
built-in in this model. Without any knowledge for the right-handed
neutrino sector, we assumed that the right-handed neutrino 
mass is proportional to a unit matrix. Modifications for
the prediction on the mixing angles are relatively minor even if
we somewhat relax this assumption, in so far as the mass hierarchy
in the Dirac mass is not disturbed. 
In conclusion, a simple Fritzsch-type model we proposed in [6] is
consistent with all existing neutrino experiments, 
but the model parameters are now restricted to a
narrow range that endows the model with a predicive power.

\newpage
\begin{table}
\hskip 0  cm
\begin{tabular}{ | c| c| c| c| c| c|} \hline
        &&   &   &    &        \\
 &$m_2/m_3$ (input)&    $m_1/m_3$  & $|U_{e3}|$   & $ |J_{CP}| $  
& $\langle m_{ee}\rangle$ (meV)\\ 
 &&   &   &  &          \\ \hline
  &&   &   &     &       \\ 
KamLAND-A \qquad 
&$0.11-0.19$& $0.007 - 0.056$ & $0.042 - 0.191$  &$\leq 0.021$ &$1.9-6.7$ \\  
 &&   &   &  &          \\\hline
  &&   &   &   &         \\
KamLAND-B\qquad 
&$0.18-0.28$&  $0.011-0.046$ & $0.081 - 0.202$& $\leq 0.025$ &$3.3-8.3$ \\ 
  &&   &   &   &         \\  \hline
  &&   &   &  &          \\
 Before KamLAND\quad 
&$0.08-0.28$& $0.005 - 0.073$ & $0.025 - 0.202$ & $\leq 0.025$ & $1.3-9.5$ \\ 
 &  &   &   &   &         \\  \hline
\end{tabular} 
\end{table} 
 Table 1 :  Summary of predictions

\newpage
{\large\bf Acknowledgements}

\vskip 1 cm
{\bf References}\par
\vskip 0.3 cm
\noindent
[1]  KamLAND Collaboration, K. Eguchi et al., hep-ex/0212021.

\noindent
[2]  SNO Collaboration: Q. R. Ahmad et al., Phys. Rev. Lett. {\bf 87} (2001)
 071301;\\ nucl-ex/0204008, 0204009.

\noindent
[3]   Super-Kamiokande Collaboration, Y.~Fukuda et al., 
 Phys.~Rev.~Lett. {\bf 81} (1998) 1562;  
J. Kameda, in {\it Proceedings of International  Cosmic Ray 
Conference, Hamburg, 2001}, edited by K. H. Kampert, G. Heinzelmann
and C. Spiering (Copernicus Gesellschaft, Katlenburg-Lindau, 2001),
Vol. 3, p.1057. 

\noindent
[4] CHOOZ Collaboration, M. Apollonio et al.,  Phys. Lett. {\bf 466B} 
  (1999) 415.

\noindent
[5] Kamiokande Collaboration, K.S. Hirata et al., 
Phys. Lett. {\bf 280B} (1992) 146.

\noindent
[6] M. Fukugita, M. Tanimoto, and T. Yanagida, Prog. Theor. Phys. {\bf 89}
 (1993) 263.

\noindent
[7] H. Fritzsch,  Phys. Lett. {\bf B73} (1978) 317; 
         Nucl. Phys. {\bf B115} (1979) 189.

\noindent
[8] For a review, see
 G. Altarelli and F. Feruglio,  hep-ph/0206077.

\noindent
[9] Z.-z. Xing,  Phys. Lett. {\bf 550B} (2002) 178.

\noindent
[10]  M. Fukugita and M. Tanimoto,  Phys. Lett. {\bf 515B} (2001) 30.

\noindent
[11]  C. Jarlskog,  Phys. Rev. Lett. {\bf 55}  (1985) 1039.

\noindent
[12]  M. Fukugita, M. Tanimoto and  T. Yanagida,  
 Phys. Rev. D{\bf 57} (1998) 4429;  Phys. Rev. D{\bf 59} (1999) 113016.

\noindent
[13] A. Zee,  Phys.~Lett. {\bf 93B} (1980) 389; {\bf B161} (1985) 141;\\
  L. Wolfenstein, Nucl. Phys. {\bf B175} (1980) 92.

\noindent
[14] S. T. Petcov,   Phys. Lett. {\bf B115} (1982) 401;\\
     C.~Jarlskog, M.~Matsuda, S.~Skadhauge and M.~Tanimoto, Phys.~Lett. 
    {\bf B449} (1999) 240;\\
     P.H.~Frampton and S.~Glashow, Phys.~Lett. {\bf B461} (1999) 95.


\newpage

\begin{figure}
\epsfxsize=12.0 cm
\vspace{-2.5 cm}
\centerline{\epsfbox{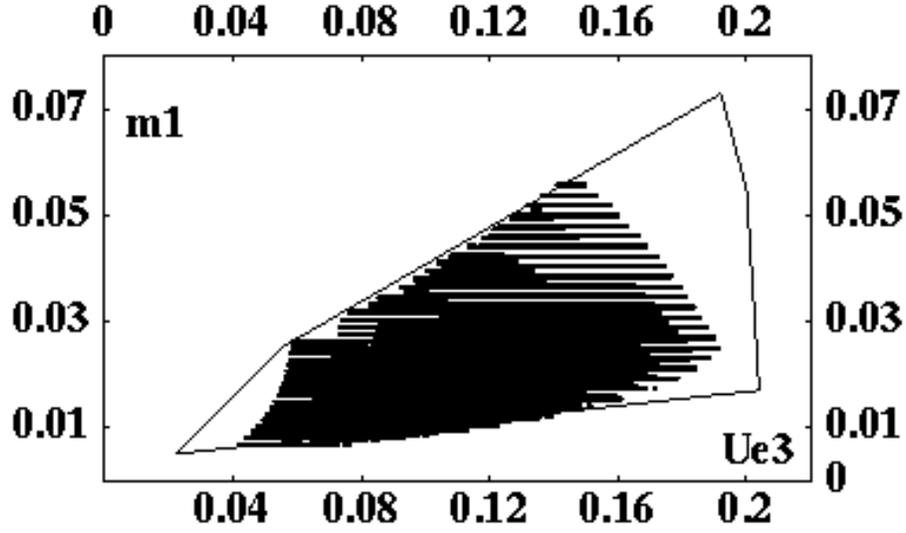}}
\vspace{-2.5 cm}
\caption{ Predicted value  of $m_1$ in the unit of $m_3$ 
 as a function of $|U_{e3}|$ in the case of KamLAND-A.  The 
outer contour shows an allowed region before the KamLAND data are available.}
\end{figure}
\begin{figure}
\epsfxsize=12.0 cm
\vspace{-2.5 cm}
\centerline{\epsfbox{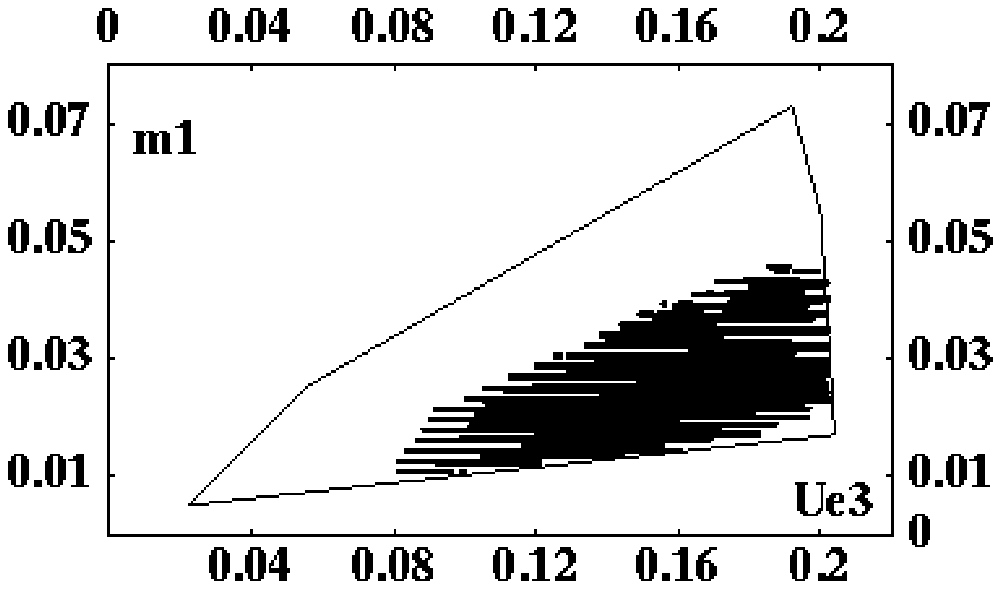}}
\vspace{-2.5 cm}
\caption{ Predicted value  of $m_1$ in the unit of $m_3$ 
 as a function of $|U_{e3}|$ in the case of KamLAND-B.  The 
outer contour shows an allowed region before the KamLAND data are available.}
\end{figure}
\begin{figure}
\epsfxsize=12.0 cm
\vspace{-2.5 cm}
\centerline{\epsfbox{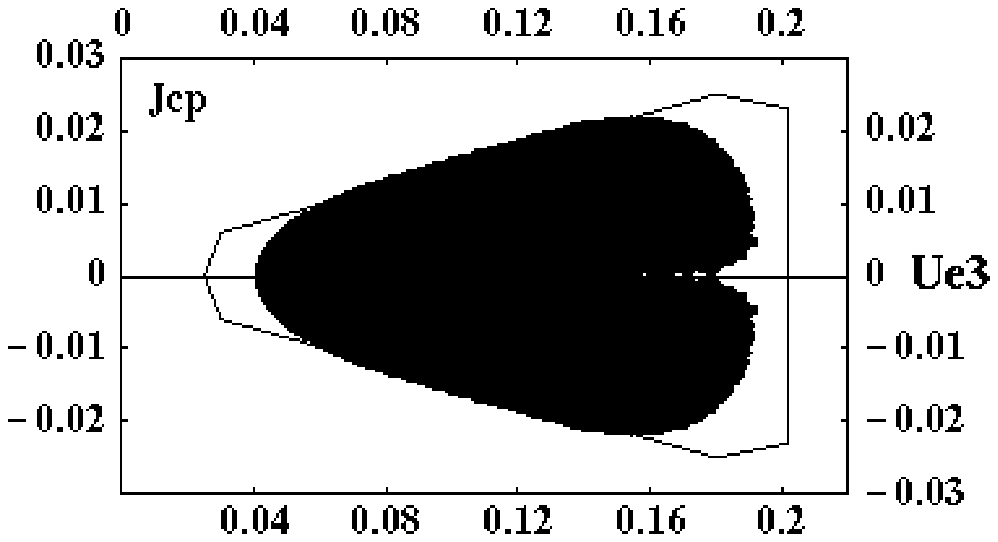}}
\vspace{-2.5 cm}
\caption{ Predicted value  of $J_{CP}$ 
 as a function of $|U_{e3}|$ in the case of KamLAND-A.  The 
outer contour shows an allowed region before the KamLAND data are available.}
\end{figure}
\begin{figure}
\epsfxsize=12.0 cm
\vspace{-2.5 cm}
\centerline{\epsfbox{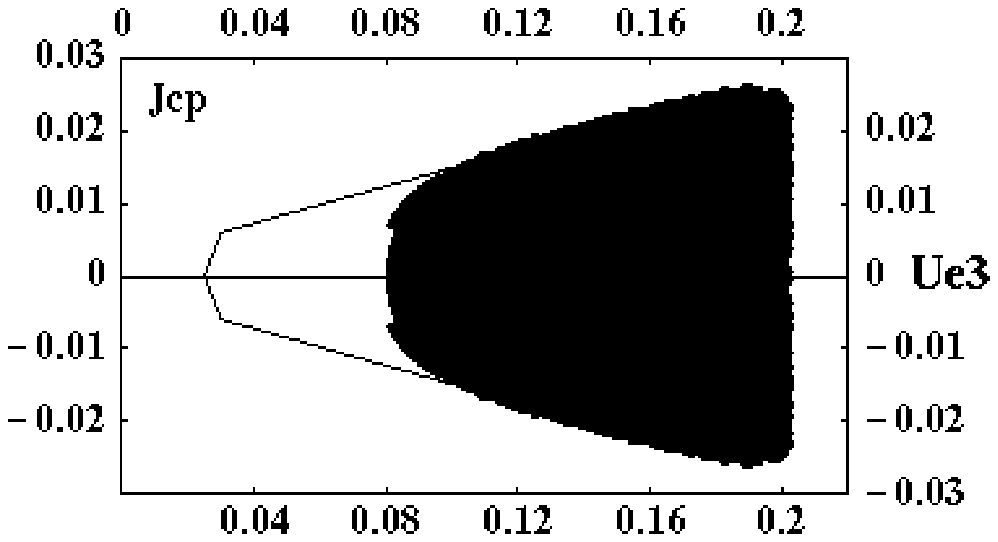}}
\vspace{-2.5 cm}
\caption{ Predicted value  of $J_{CP}$ 
 as a function of $|U_{e3}|$ in the case of KamLAND-B.  The 
outer contour shows an allowed region before the KamLAND data are available.}
\end{figure}
\begin{figure}
\epsfxsize=12.0 cm
\vspace{-2.5 cm}
\centerline{\epsfbox{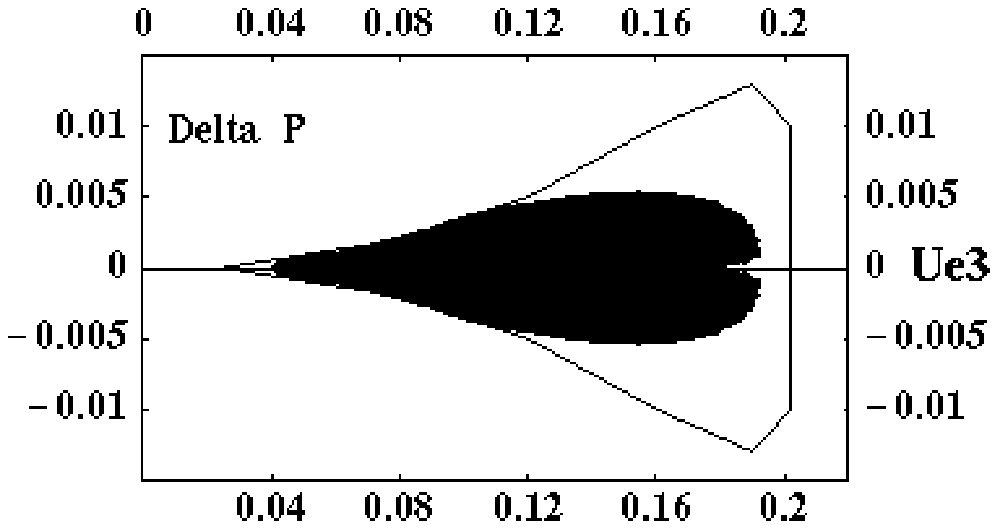}}
\vspace{-2.5 cm}
\caption{ Predicted value  of $\Delta P$ 
 as a function of $|U_{e3}|$ in the case of KamLAND-A. The 
outer contour shows an allowed region before the KamLAND data are available. }
\end{figure}
\begin{figure}
\epsfxsize=12.0 cm
\vspace{-2.5 cm}
\centerline{\epsfbox{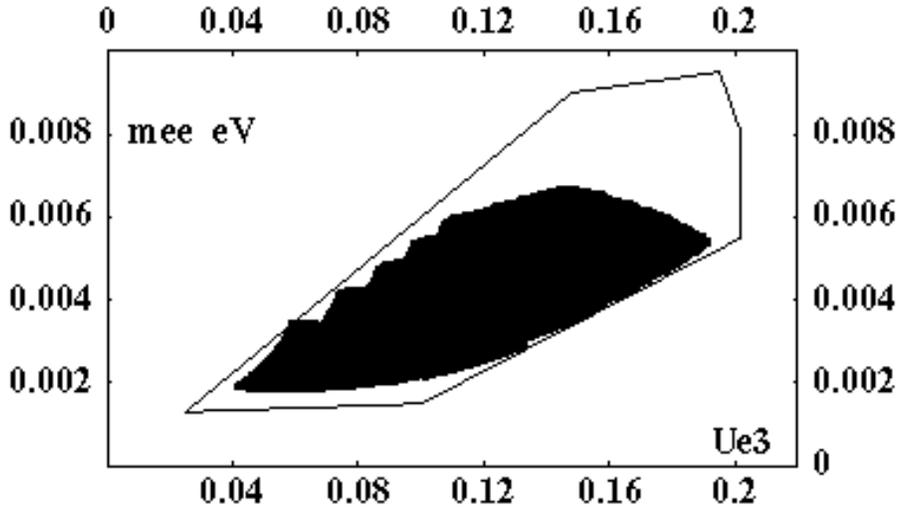}}
\vspace{-2.5 cm}
\caption{ Predicted value  of  $\langle m_{ee}\rangle$ (eV) 
 as a function of $|U_{e3}|$ in the case of KamLAND-A. The 
outer contour shows an allowed region before the KamLAND data are available. }
\end{figure}
\end{document}